\documentclass[aip,apl,amsmath,amssymb, reprint]{revtex4-1}
\usepackage[utf8]{inputenc}
\usepackage[T1]{fontenc}
\usepackage{graphicx}
\usepackage{hyperref}
\begin{document}

\title{A long-lived capacitively shunted flux qubit embedded in a 3D cavity} 

\author{Leonid V. Abdurakhimov}
\email{leonid.abdurakhimov.nz@hco.ntt.co.jp}

\author{Imran Mahboob}
\author{Hiraku Toida}
\author{Kousuke Kakuyanagi}
\author{Shiro Saito}
\affiliation{NTT Basic Research Laboratories, NTT Corporation, 3-1 Morinosato-Wakamiya, Atsugi, Kanagawa 243-0198, Japan}

\date{\today}

\begin{abstract}
We report the experimental realization of a 3D capacitively-shunt superconducting flux qubit with long coherence times. At the optimal flux bias point, the qubit demonstrates energy relaxation times in the 60--90\,$\mu$s range, and Hahn-echo coherence time of about 80\,$\mu$s which can be further improved by dynamical decoupling. Qubit energy relaxation can be attributed to quasiparticle tunneling, while qubit dephasing is caused by flux noise away from the optimal point. Our results show that 3D c-shunt flux qubits demonstrate improved performance over other types of flux qubits which is advantageous for applications such as quantum magnetometry and spin sensing.
\end{abstract}

\maketitle 

Improving the performance of superconducting flux qubits is crucial for the development of emerging quantum technologies, such as quantum annealing\,\cite{Johnson2011}, quantum magnetometry\,\cite{Bal2012}, and spin sensing\,\cite{Toida2019}. A conventional flux qubit consists of a superconducting loop interrupted by three Josephson junctions, and transition frequencies between qubit states can be controlled by applying external magnetic flux through the qubit loop\,\cite{Mooij1999,Orlando1999}. The decoherence of qubit states is caused by unwanted interactions with the environment, and different techniques have been employed to mitigate their effects. 
One of the major sources of the decoherence of a conventional flux qubit is magnetic flux noise, and it was shown theoretically that susceptibility of a flux qubit to the flux noise (as well as charge noise) can be significantly reduced by shunting the smaller Josephshon junction by an additional capacitance\,\cite{You2007}. Initial experimental studies of c-shunt flux qubits coupled to a coplanar resonator (2D c-shunt flux qubits) reported a modest improvement of coherence times\,\cite{Steffen2010,Corcoles2011}, but later experiments demonstrated energy relaxation times $T_1$ of up to 55\,$\mu$s\,\cite{Yan2016} and spin-echo decoherence times $T_{2\mathrm{E}}$ of about 80\,$\mu$s\,\cite{Yan2018}. 
Besides magnetic flux noise, another possible qubit decoherence mechanism is the unintended interaction between a qubit and spurious microwave modes. As originally demonstrated in experiments with transmon qubits\,\cite{Paik2011,Rigetti2012}, it is possible to create a well-controlled electromagnetic environment for a qubit by coupling it to a 3D microwave cavity. Moreover, it was found that surface dielectric losses can be reduced by using 3D qubit designs\,\cite{Wang2015}. In the case of flux qubits, it was shown that, by embedding a conventional flux qubit in a 3D cavity, intrinsic energy relaxation time can reach 20\,$\mu$s and pure dephasing times can be up to 10\,$\mu$s\,\cite{Stern2014}.
In this work, we combined the above approaches in order to reduce qubit decoherence due to magnetic flux noise, charge noise and coupling to higher microwave modes.

Here we present the realization of a c-shunt flux qubit coupled to a 3D microwave cavity. At the optimal working point of magnetic flux bias, we observed energy-relaxation times $T_1$ in the range of 60--90\,$\mu$s, and spin-echo coherence times $T_{2\mathrm{E}}$ of up to 80\,$\mu$s. Using a dynamical decoupling Carr-Purcell-Meiboom-Gill (CPMG) sequence, it was possible to reach coherence times of up to 160\,$\mu$s. Thus, 3D c-shunt flux qubits demonstrate excellent coherence times, which are comparable or exceed the best values reported for other types of flux qubits. 

\begin{figure}
    \centering
    \includegraphics{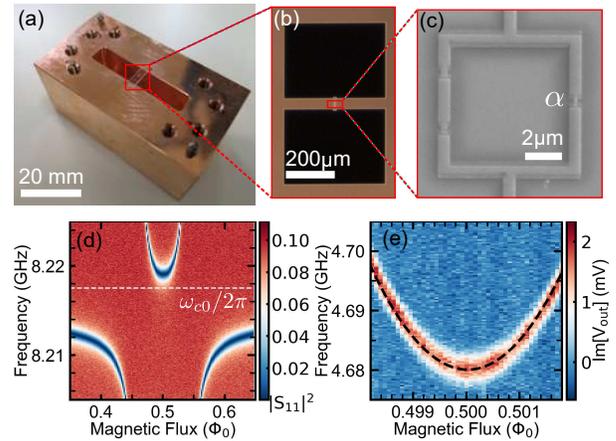}
    \caption{(a) A photograph of the c-shunt flux qubit mounted inside the 3D cavity. (b) A microscope image of the qubit. The shunt capacitance of the qubit was formed by two large rectangular pads. (c) An SEM micrograph of the qubit loop interrupted by three Josephson junctions. (d) Microwave response of the 3D cavity coupled to the qubit. Microwave probe power at the cavity input port was about -135\,dBm. The dashed line corresponds to the bare cavity frequency $\omega_{\mathrm{c}0}/2\pi$, obtained in a separate high-microwave-power measurement. (e) Results of the microwave spectroscopy of the qubit transition from the ground state to the excited state using dispersive readout. The dashed line corresponds to the fitting using analytical equations described in the text.}
    \label{fig1}
\end{figure}

\begin{figure*}
    \centering
    \includegraphics{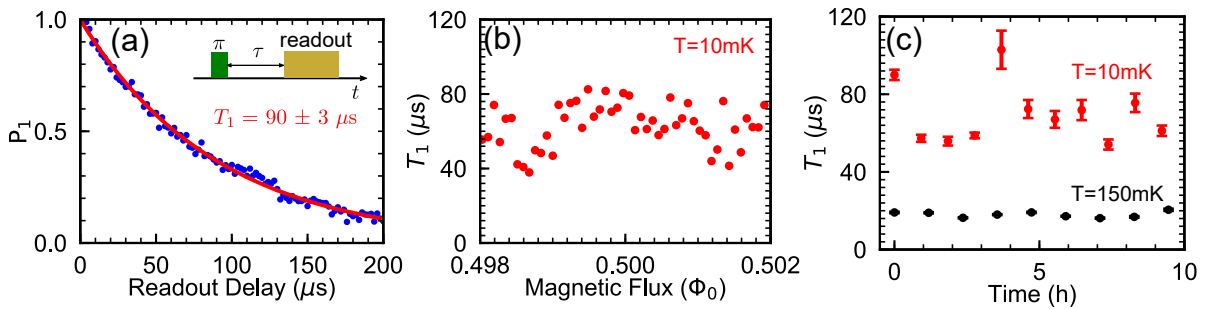}
    \caption{Energy-relaxation time $T_1$ of the 3D c-shunt flux qubit. (a) Typical trace of the inversion recovery signal at the optimal point. Dots correspond to the experimental data, the red line is the exponential fitting with $T_1\approx90$\,$\mu$s. Inset: a schematic of the measurement pulse sequence. (b) Typical dependence of $T_1$ on the magnetic flux at $T=10$\,mK. (c) Temporal stability of $T_1$ at $T=10$\,mK and $T=150$\,mK at the optimal flux bias point.}
    \label{fig2}
\end{figure*}

Experiments were performed in a dilution refrigerator with a base temperature of 10\,mK\,\cite{SM}. A c-shunt flux qubit was mounted in the center of a copper 3D microwave cavity\,[Figs.\,\ref{fig1}(a-c)]. The qubit was fabricated on a high-quality single-crystal C-plane sapphire substrate and consisted of a superconducting aluminium square loop interrupted by three Josephson junctions\,[Fig.\,\ref{fig1}(c)]. Two junctions were almost identical and had similar values of the Josephson energy $E_J$ and the charging energy $E_\mathrm{C} = e^2/2C_\mathrm{J}$. The third junction, whose area was made smaller by a factor of $\alpha < 0.5$, was shunted by a large coplanar capacitance $C_\mathrm{S} \gg C_\mathrm{J}$ formed by two rectangular aluminium pads\,[Fig.\,\ref{fig1}(b)], which also served as a dipole antenna for the coupling of the qubit to the cavity. The qubit chip was glued to the cavity by GE Varnish in order to maintain good thermal contact. To apply magnetic flux bias to the qubit, a small custom-made solenoid magnet was attached outside the 3D cavity. The cavity was attached to a cold finger of the dilution refrigerator and surrounded by a 3-layer magnetic shield consisting of two mu-metal layers and a copper-coated aluminium layer. Custom-made Eccosorb microwave filters were used to suppress quasiparticle generation by microwave photons with frequencies $\hbar \omega > 2\Delta_0$, where $\Delta_0 \approx 200$\,$\mu$eV is the superconducting gap of a thin aluminium film. To suppress qubit dephasing by the residual thermal photons in the cavity\,\cite{Yan2018}, a sufficient number of microwave cryogenic attenuators were used on the input line, and several microwave cryogenic isolators were installed on the output line\,\cite{SM}.

The qubit was coupled to the lowest frequency TE101 mode of the 3D cavity. By measuring $S_{11}$ parameter of the cavity using a vector network analyzer, anticrossing behaviour between the cavity mode and the qubit was observed at low microwave-probe power\,[Fig.\,\ref{fig1}(d)]. At the optimal working point --- where the applied magnetic flux $\Phi$ was equal to a half of the magnetic flux quantum $\Phi_0=h/2e$ --- the frequency of dressed-state cavity mode $\omega_{c}/2\pi$ was 8.219\,GHz. At high microwave-probe power, the qubit and the cavity were decoupled\,\cite{Bishop2010,Boissonneault2010}, and the bare cavity frequency was found to be $\omega_{\mathrm{c}0}/2\pi \approx$\,8.2175\,GHz. The cavity was slightly undercoupled with the loss rate due to external coupling of $\kappa_\mathrm{c} \approx 0.6$\,MHz, and the internal loss rate of $\kappa_\mathrm{i} \approx 0.7$\,MHz. The total linewidth of the cavity mode was $\kappa \approx 1.3$\,MHz which corresponded to a quality factor of $Q \approx 6000$.

Qubit measurements were performed using cavity QED architecture in the dispersive regime\,\cite{Koch2007}, when the detunings between the qubit transitions $\omega_{ij}$ and the cavity mode $\omega_\mathrm{c}$ are larger than the coupling between them. When the qubit was excited from the ground $i=0$ to the first excited state $j=1$, the cavity frequency was shifted by the cavity pull $2 \chi /2\pi \approx 1.8$\,MHz. By fixing the frequency of the cavity readout pulse at the value corresponding to the ground state, the qubit state was determined by measuring the amplitude of the reflected microwave signal.  As shown in Fig.\,\ref{fig1}(e), the frequency of the qubit transition between the ground and excited state was  $\omega_{01}/2\pi\approx 4.68$\,GHz at the optimal point. The anharmonicity $A = \omega_{12} - \omega_{01}$ was found to be $A/2\pi \approx780$\,MHz in a separate measurement of the $\omega_{12}$ qubit transition using two-tone qubit microwave probing. According to the theoretical model of the coupling between a qubit and a cavity in the dispersive limit\,\cite{Koch2007}, cavity frequency shifts are given by
\begin{align}
    \chi=\chi_{01}-\chi_{12}/2, \\
    \omega_{\mathrm{c}0} - \omega_\mathrm{c} = \chi_{01},
\end{align}
where $\chi_{ij}=g_{ij}^2/(\omega_{ij}-\omega_{\mathrm{c}0})$ with $g_{ij}$ corresponding to the coupling strength between the cavity mode and the qubit transition between $\left|i\right>$ and $\left|j\right>$ states. From the experimental data, $g_{01} / 2\pi \approx 73$\,MHz and $g_{12}/ 2\pi \approx 115$\,MHz were extracted.

\begin{figure*}
    \centering
    \includegraphics{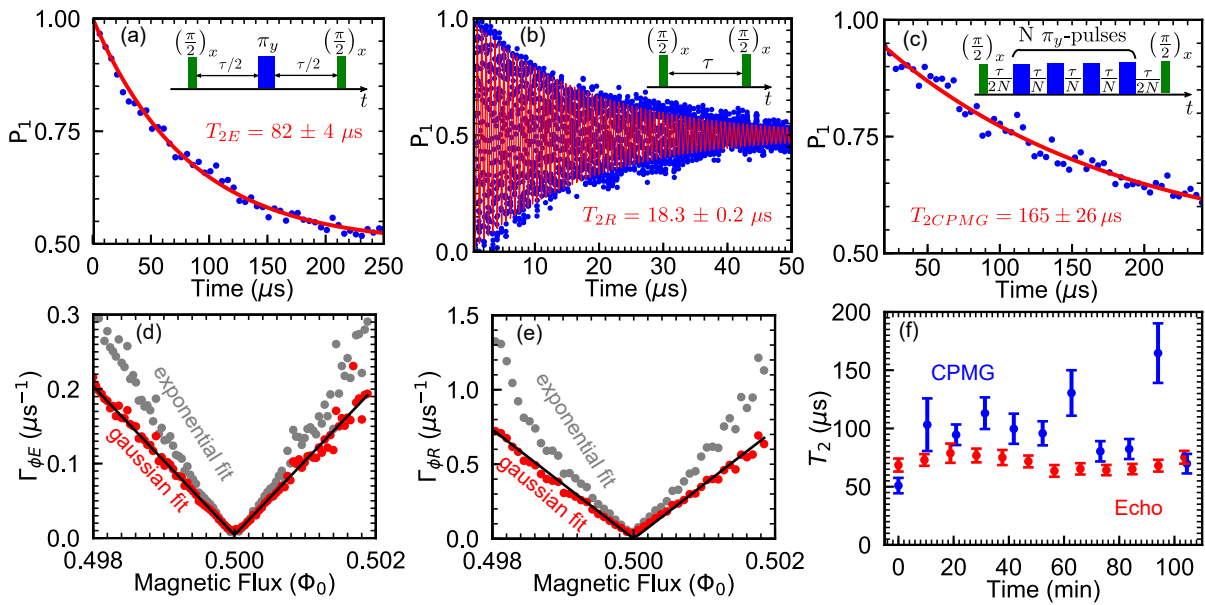}
    \caption{Coherence times of the 3D c-shunt flux qubit at $T=10$\,mK. (a), (b), and (c) Typical traces for Hahn-echo decay, Ramsey fringes and CPMG ($N=20$) decay signals, respectively, at the optimal point. Blue dots are experimental data points, red lines are fitting curves with the given characteristic relaxation times. Insets: schematics of the corresponding measurement pulse sequences. (d,e) The pure dephasing rates $\Gamma_{\phi E}$ and $\Gamma_{\phi R}$ derived from the Hahn-echo and Ramsey decay experimental data, respectively, as a function of the applied magnetic flux. As described in the text, dephasing rates were calculated using Gaussian fitting (red points) or pure exponential fitting (gray points), and the former can be fitted, in turn, by a linear function (black straight lines) away from the optimal point. (f) Temporal variations of the dephasing time at the optimal point obtained in Hahn-echo measurements (red points) and the measurements with a dynamical decoupling (blue points).}
    \label{fig3}
\end{figure*}

In the case of a large shunting capacitance  $C_\mathrm{S} \gg C_\mathrm{J}$, the Hamiltonian of a c-shunt flux qubit can be reduced to a one-dimensional form\,\cite{Steffen2010} and treated by perturbation theory\,\cite{SM}. The qubit spectrum can then be described as\,\cite{SM}
\begin{equation}
 \hbar \omega_{01} = \Delta + 2 \varepsilon^2/\Delta,   
\end{equation}
where the energy gap $\Delta$ and the flux-dependent energy shift $\varepsilon$ are defined by
\begin{align}
\Delta = \sqrt{4 E_{C_\mathrm{S}} E_\mathrm{J} (1-2\alpha)} +  \frac{8\alpha-1}{4(1-2\alpha)} E_{C_\mathrm{S}}, \\   
\varepsilon = 2 \sqrt{2} \pi \alpha E_\mathrm{J} \left( \frac{E_{C_\mathrm{S}}}{E_\mathrm{J}(1-2\alpha)}\right)^{1/4}\left(\frac{\Phi}{\Phi_0}-0.5\right).
\end{align}
Here, $E_{C_\mathrm{S}} = e^2/2C_\mathrm{S}$ is the effective charging energy determined by the shunt capacitance $C_\mathrm{S}$. By fitting the experimental data using these equations, the qubit parameters $\alpha\approx0.41$, $C_\mathrm{S}\approx78$\,fF ($E_{C_\mathrm{S}}\approx 0.25$\,GHz), and $E_\mathrm{J}\approx85$\,GHz were obtained. The spectrum of the c-shunt flux qubit can also be reproduced by solving numerically the full two-dimensional Hamiltonian of a c-shunt flux qubit given in the Ref.\,\onlinecite{You2007} in the phase basis on a grid of $80\times80$ with the following parameters: $\alpha=0.437$, $C_\mathrm{S}=60$\,fF, $E_\mathrm{J}=136.75$\,GHz, and $E_\mathrm{C}\approx3.2$\,GHz ($C_\mathrm{J}=6$\,fF). It should be noted that --- since the parameter $\alpha$ was less than 0.5 --- the potential energy of the qubit used in this work had only one minimum, while the potential energy of a conventional flux qubit has a double-well structure. Nevertheless, the wording ``c-shunt flux qubit'' is employed in order to be consistent with the literature on c-shunt flux qubits where similar values of $\alpha$ are used\,\cite{Steffen2010,Yan2016}.

The 3D c-shunt flux qubit demonstrated long energy-relaxation times $T_1$ in the range of 60--90\,$\mu$s at a temperature of 10\,mK, as shown in Fig.\,\ref{fig2}. The relaxation time was roughly independent of the applied magnetic flux in the vicinity of the optimal point of 0.5\,$\Phi_0$ [Fig.\,\ref{fig2}(b)], except for two values, located symmetrically about the optimal point, where $T_1$ decreased to about $40$\,$\mu$s, and which can be attributed to weak coupling of the qubit to a resonant two-level system\,\cite{Stern2014,Barends2013}. At $T=10$\,mK, the qubit demonstrated variations in relaxation times under repeated measurements, as shown in Fig.\,\ref{fig2}(c). At a temperature of 150\,mK, the $T_1$ relaxation time was stable around 20\,$\mu$s [Fig.\ref{fig2}(c)]. The temperature dependence of $T_1$ can be ascribed to quasiparticle tunneling through a Josephson junction\,\cite{Catelani2011a, Catelani2011b}. Indeed, an upper limit of the density of non-equilibrium quasiparticles $n_{qp}\leq0.6$\,$\mu$m$^{-3}$ can be estimated\,\cite{SM} which is comparable to the values obtained in experiments with other types of superconducting qubits coupled to a 3D cavity\,\cite{Paik2011,Stern2014}. Consequently, the temporal variations of the $T_1$ time at $T=10$\,mK were likely related to quasiparticle fluctuations. However, it should be noted that other possible noise sources, including charge noise and magnetic flux noise, could not be completely excluded\,\cite{Yan2016,Muller2015}. Relaxation due to Purcell effect was estimated to be negligible with the Purcell-limited relaxation time of $T_{1P} = (\kappa \times g_{01}^2/(\omega_{01}-\omega_\mathrm{c})^2)^{-1}\approx 1.6$\,ms.

The qubit coherence times were measured using Hahn-echo, Ramsey and CPMG pulse sequences [Fig.\,\ref{fig3}], and, using exponential fitting functions for the signal envelopes, $T_{2\mathrm{E}}\approx 80$\,$\mu$s, $T_{2\mathrm{R}}\approx 18$\,$\mu$s and $T_{2\mathrm{CPMG}}$ up to 160\,$\mu$s were obtained, respectively, at the optimal point when $T=10$\,mK\,[Figs.\,\ref{fig3}(a-c)]. At $T=150$\,mK, Hahn-echo and Ramsey coherence times of $T_{2\mathrm{E}}\approx2$\,$\mu$s and $T_{2\mathrm{R}}\approx1.5$\,$\mu$s were measured, respectively.

Next, the dependence of qubit dephasing rates on the applied magnetic flux was investigated\,[Figs.\,\ref{fig3}(d,e)]. Away from the optimal point, the qubit dephasing was caused by $1/\omega$ flux noise.   Assuming that the spectral density of $1/\omega$ noise is defined as $S(\omega)=A_\Phi/\omega$ in the frequency range limited by an infrared cutoff frequency $\omega_{\mathrm{ir}}$ and an ultraviolet cutoff frequency, one can show\,\cite{Ithier2005} that Hahn-echo and Ramsey decays can be described by Gaussian functions
\begin{equation}
    F_{\mathrm{E},\mathrm{R}}=e^{-t/2T_1}e^{-(\Gamma_{\phi \mathrm{E},\mathrm{R}}t)^2},
    \label{eq:gauss}
\end{equation}
with pure dephasing rates $\Gamma_{\phi \mathrm{E}} = \sqrt{A_\Phi \ln{2}} |\partial \omega_{01} / \partial f|$ and $\Gamma_{\phi \mathrm{R}} = \sqrt{A_\Phi \ln{(1/\omega_{\mathrm{ir}}t)}} |\partial \omega_{01} / \partial f|$, respectively, where $f$ is the normalized magnetic flux, $f=\Phi / \Phi_0$. Due to the low signal-to-noise ratio of the raw experimental data, it was not possible to reliably distinguish between a Gaussian decay described by Eq.\,(\ref{eq:gauss}) and a pure exponential decay. Therefore, two values for dephasing rates $\Gamma_{\phi \mathrm{E},\mathrm{R}}$ and $\tilde{\Gamma}_{\phi \mathrm{E},\mathrm{R}}$ were extracted for each trace by using Eq.\,(\ref{eq:gauss}) (red points in Figs.\,\ref{fig3}(d,e)) and a pure exponential decay function $G_{\mathrm{E},\mathrm{R}}=e^{-t/2T_1}e^{-\tilde{\Gamma}_{\phi \mathrm{E},\mathrm{R}}t}$ (gray points in Figs.\,\ref{fig3}(d,e)), respectively. Since only $\Gamma_{\phi \mathrm{E},\mathrm{R}}$ could be reproduced consistently by a linear function away from the optimal point, one can conclude that the experimental data was best fit by Eq.\,(\ref{eq:gauss}), and the qubit dephasing was indeed caused by $1/\omega$ flux noise. From the $\Gamma_{\phi \mathrm{E}}$ fitting, the amplitude of the flux noise was found to be $A_\Phi\approx(1.8\mu \Phi_0)^2$ which is comparable to the values reported in the literature\,\cite{Yoshihara2006, Bylander2011, Stern2014}.
Far from the optimal point, the experimental value of the ratio $\Gamma_{\phi \mathrm{R}} / \Gamma_{\phi \mathrm{E}}\approx 3.7$ was close to the theoretically estimated value $\Gamma_{\phi \mathrm{R}} / \Gamma_{\phi \mathrm{E}} = \sqrt{\ln{(1/\omega_{\mathrm{ir}}t)}/\ln{2}}\lesssim 5$. The theoretical value was calculated using $\omega_{\mathrm{ir}}/2\pi\approx 1/\Delta t$ and $t\approx T_{2\mathrm{R}}$, where $\Delta t\approx2.45$\,s was the acquisition time of a single data point, and $T_{2\mathrm{R}} \approx 1$\,$\mu$s was taken far from the optimal point. 

Similar to other Josephson qubits\,\cite{Koch2007,Bylander2011}, dephasing of the 3D c-shunt flux qubit at the optimal point was probably caused by the charge noise or critical-current fluctuations. The dephasing time due to the thermal-photon noise was estimated\,\cite{SM} to be $T_{\phi}^{th}\approx 3$\,ms, and hence it could not explain the observed value of $T_{\phi}= (T_{2E}^{-1} - (2 T_1)^{-1})^{-1}\approx 160$\,$\mu$s. By using the CPMG dynamical decoupling sequence with $N=20$ $\pi$-pulses, the $2T_1$ limit for the decoherence time was reached\,[Figs.\,\ref{fig3}(c,f)], with the temporal variation of $T_{2\mathrm{CPMG}}$ time most likely stemming from the temporal variation of the $T_1$ time as detailed in Fig.\,\ref{fig2}(c). Using filter function formalism\,\cite{Biercuk2009,Bylander2011}, it can be shown that the application of the CPMG sequence effectively results in the filtering out of the low-frequency part of the dephasing noise\,\cite{SM}. Therefore, the improvement of the coherence time in the CPMG experiment is consistent with the assumption that the qubit decoherence in the Ramsey and Hahn-echo measurements was caused by $1/\omega$ charge or critical-current noises. 

In conclusion, we realized a 3D c-shunt flux qubit and investigated its relaxation mechanisms. The values obtained for the relaxation times are comparable or exceed the best reported values for flux qubits\,\cite{Stern2014, Yan2016, Yan2018}. The qubit energy relaxation can be attributed to quasiparticle tunneling, while its dephasing is caused by charge noise or critical-current fluctuations at the optimal point and $1/\omega$ flux noise elsewhere. Flux sensitivity of the given qubit can reach\,\cite{note1} a value of  $S_\Phi\approx1.2\times10^{-8}$\,$\Phi_0/\sqrt{\mathrm{Hz}}$, or approximately 0.5\,pT/$\sqrt{\mathrm{Hz}}$ in terms of magnetic field sensitivity, thus indicating the prospect of significant improvement over previous works\,\cite{Danilin2018}. Therefore, the usage of such 3D c-shunt flux qubits will be highly beneficial for applications in quantum magnetometry and spin sensing. 

\begin{acknowledgments}
The authors are grateful to Prof. Y. Nakamura of the University of Tokyo and RIKEN for supporting the initial development of this project. This work was partially supported by CREST (JPMJCR1774), JST.
\end{acknowledgments}

\providecommand{\noopsort}[1]{}\providecommand{\singleletter}[1]{#1}%
%

\widetext
\clearpage
\begin{center}
\textbf{\large Supplemental Materials \\ ``A long-lived capacitively shunted flux qubit embedded in a 3D cavity''}
\end{center}
\setcounter{equation}{0}
\setcounter{figure}{0}
\setcounter{table}{0}
\setcounter{page}{1}
\makeatletter
\renewcommand{\theequation}{S\arabic{equation}}
\renewcommand{\thefigure}{S\arabic{figure}}
\renewcommand{\bibnumfmt}[1]{[S#1]}
\renewcommand{\citenumfont}[1]{S#1}

\section{Experimental setup}
\begin{figure}[hb]
    \centering
    \includegraphics[width=0.7\textwidth]{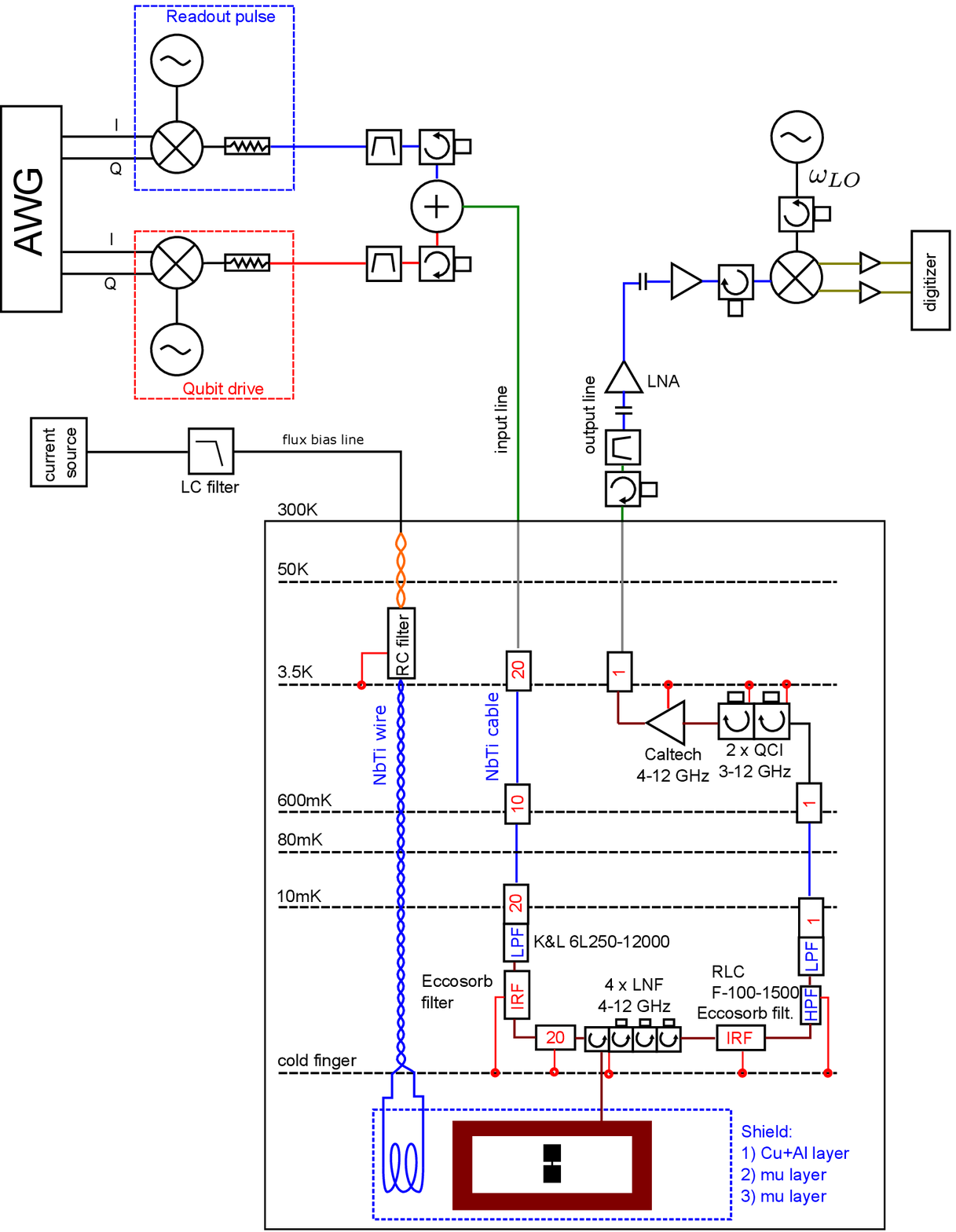}
    \caption{A schematic of the experimental setup employed to measure the c-shunted flux qubit.}
    \label{fig1S}
\end{figure}

The scheme of the experimental setup is shown in the Fig.\,\ref{fig1S}. Measurements were performed using Labber software\,\cite{Labber}.
The 3D cavity was fabricated from OFHC copper and it had internal dimensions of 5\,mm $\times$ 27\,mm $\times$ 24\,mm. The cavity body consisted of two parts enabling access to the cavity volume, and two small depressions were made on the opposite side walls of the cavity bottom part to accommodate the qubit. An SMA connector was mounted on the cavity side wall with the central pin protruding into the cavity through a small hole in the wall. It was possible to adjust the coupling between the cavity and the microwave line by changing the length of the central pin projecting out of the internal side wall.

The qubit was fabricated on a sapphire substrate with dimensions of 2\,mm $\times$ 7\,mm $\times$ 0.5\,mm and it was annealed before deposition of the shunt-capacitor aluminium layer using an e-beam evaporator. The dimensions of the capacitor pads were 300$\mu$m $\times$ 400$\mu$m, and the qubit loop was approximately 7$\mu$m $\times$ 7$\mu$m. Josephson junctions were fabricated by double-angle shadow evaporation of aluminium using a Plassys system.

Microwave pulses for the qubit drive and cavity readout were generated using an arbitrary waveform generator Tektronix AWG5014C, a 4-channel scalar microwave source Anapico APMS20G-4, and MarkiMicrowave IQ mixers. Gaussian pulses were used for the qubit drive with $4\sigma$ truncation. To avoid issues related to LO leakage of the IQ mixer, the envelopes of qubit drive I/Q pulses were modulated with a frequency of 123\,MHz for single-sideband upconversion. Microwave power for qubit drive pulses was relatively low with a typical duration for the $\pi/2$ pulse of $\sigma \approx 70$\,ns. The cavity readout pulse was typically several microseconds long. Repetition times for the pulse sequences were chosen to be at least $3\times T_1$. Inside the dilution refrigerator, microwave lines were carefully filtered, attenuated/isolated as described in the main text. The output signal was amplified by a cryogenic Caltech amplifier mounted at the 4K stage, and by an LNA amplifier at room temperature. The detection was performed using a heterodyne scheme with the modulation frequency of 29\,MHz. The downconverted output signal was digitized and averaged by an Acqiris U1084A card and demodulated by software means.

A custom-made flux bias coil was powered by a current source ADCMT 6166 with the supply lines filtered by a MiniCircuits LC filter at room temperature and an Aivon RC filter at the 4K stage.

\section{Perturbation theory for the c-shunt flux qubit\label{section:perturbation}}

Assuming $E_J/E_c \gg 1$ and $C_S \gg C$, analytical equations for the qubit transition frequency $\omega_{10}$ and anharmonicity $A$ at the optimal working point can be derived using the perturbation approach described in\,\cite{Koch2007s}. Here the relevant derivations are briefly outlined using common notations (for more details, see the references\,\cite{You2007s, Koch2007s}).

The Hamiltonian of a c-shunt flux qubit is given by\,\cite{You2007s}:
\begin{equation}
    H = K + U,
\end{equation}
with the potential energy
\begin{equation}
U = 2E_J (1-\cos{\varphi_p}\cos{\varphi_m}) + \alpha E_J (1-\cos{(2\pi f + 2\varphi_m)}),
\label{eq:potential}
\end{equation}
and the kinetic energy
\begin{equation}
    K = E_p n_p^2 + E_m n_m^2.
\end{equation}
Assuming $C_S \gg C$ and following the work\,\cite{Steffen2010s}, the Hamiltonian at the optimal point ($f=\Phi / \Phi_0 =0.5$) is obtained by neglecting terms with $\varphi_p$:
\begin{equation}
    H = \frac{e^2}{2C_S} n^2 + 2E_J(1-\cos{\varphi})+\alpha E_J (1+\cos{2\varphi}),
\end{equation}
where $n\equiv n_m = - i \partial / \partial \varphi_m$ and $\varphi \equiv \varphi_m$.
By defining $E_{C_S} = e^2/2C_S$ and using the cosine expansion, the Hamiltonian can be expressed as:
\begin{equation}
    H = E_{C_S} n^2 + 2\alpha E_J + E_J(1-2\alpha)\varphi^2 + E_J \frac{8\alpha-1}{12}\varphi^4.
    \label{eq:fourth-order}
\end{equation}
Introducing the operator $\tilde n=\hbar n$, equation\,(\ref{eq:fourth-order}) is re-written as:
\begin{equation}
    H = \frac{E_{C_S}}{\hbar^2} {\tilde n}^2 + 2\alpha E_J + E_J(1-2\alpha)\varphi^2 + E_J \frac{8\alpha-1}{12}\varphi^4.
    \label{eq:effective-hamiltonian}
\end{equation}
The commutation relation between the operators $\varphi$ and $\tilde n$ is $[\varphi,\tilde n]=i\hbar$, and equation\,(\ref{eq:effective-hamiltonian}) is similar to the Hamiltonian of a quantum harmonic oscillator $H_{OSC}=\frac{p^2}{2m}+\frac{1}{2}m\omega^2x^2$ with the $x^4$-term-like perturbation. Following the standard approach, the annihilation and creation operators are expressed as:
\begin{equation}
    a = \frac{1}{\sqrt{2}}\left(\ \frac{E_J(1-2\alpha)}{E_{C_S}}\right)^{1/4} \left( \varphi + i \sqrt{\frac{E_{C_S}}{E_J(1-2\alpha)}}n\right),
\end{equation}
and
\begin{equation}
    a^\dagger = \frac{1}{\sqrt{2}}\left(\ \frac{E_J(1-2\alpha)}{E_{C_S}}\right)^{1/4} \left( \varphi - i \sqrt{\frac{E_{C_S}}{E_J(1-2\alpha)}}n\right).
\end{equation}
The Hamiltonian\,(\ref{eq:effective-hamiltonian}) can then be written as
\begin{equation}
    H = \sqrt{4 E_{C_S} E_J (1-2\alpha)} \left( a^\dagger a + \frac{1}{2}\right) + 2\alpha E_J + \frac{1}{48}\frac{8\alpha-1}{1-2\alpha}E_{C_S}\left( a^\dagger + a \right)^4.
\end{equation}
Using perturbation theory, the equation for the eigenenergies is obtained:
\begin{equation}
    E_m = \sqrt{4 E_{C_S} E_J (1-2\alpha)} \left( m + \frac{1}{2}\right) + 2\alpha E_J + \frac{1}{48}\frac{8\alpha-1}{1-2\alpha}E_{C_S}\left( 6m^2 +6m +3 \right).
    \label{eq:eigenenergy}
\end{equation}
From equation\,(\ref{eq:eigenenergy}), the qubit transition frequency $\omega_{10}$ and anharmonicity $A$ at the optimal point are determined as:
\begin{equation}
    \Delta \equiv E_1 - E_0 = \hbar \omega_{10} \bigg|_{f=f_0} =  \sqrt{4 E_{C_S} E_J (1-2\alpha)} + \frac{8\alpha-1}{4(1-2\alpha)}E_{C_S},
    \label{eq:freq}
\end{equation}
and, respectively,
\begin{equation}
    \hbar A = (E_2 - E_1) - (E_1 - E_0) = \frac{8\alpha-1}{4(1-2\alpha)}E_{C_S}.
    \label{eq:anharmonicity}
\end{equation}
It should be noted that Eqs.\,(\ref{eq:freq},\ref{eq:anharmonicity}) are consistent with equations described in the Supplementary Materials of Ref.\,\cite{Yan2016s}.

This formalism can also enable the dependence of the qubit transition frequency on the applied magnetic flux to be derived by expanding the potential energy (see Eq.\,(\ref{eq:potential})) near $f_0=0.5$ in the following form:

\begin{equation}
    U \approx U_0 + (f-f_0) \cdot \frac{\partial U}{\partial f} \bigg|_{f=f_0} = U_0 + U_1,
\end{equation}
where
\begin{equation}
    U_1 = - 2\pi\alpha E_J \sin{[2\varphi_m]} \cdot (f-f_0) \approx - 4 \pi \alpha E_J \varphi_m \cdot (f-f_0).
\end{equation}
Then, the transition frequency can be expressed as: 
\begin{equation}
 \hbar \omega_{10} \approx \Delta + \frac{2 \varepsilon^2}{\Delta},
 \label{eq:spectrum}
\end{equation}
where
\begin{equation}
 \varepsilon \equiv \left|\left<1|U_1|0\right>\right|  = 2 \sqrt{2} \pi \alpha E_J \left( \frac{E_{C_S}}{E_J(1-2\alpha)}\right)^{1/4}\left(\frac{\Phi}{\Phi_0}-0.5\right). 
 \label{eq:delta}
\end{equation}
By using Eq. (\ref{eq:freq}), (\ref{eq:spectrum}), and (\ref{eq:delta}) to fit the experimental data, the values $\alpha\approx0.41$, $C_S \approx 78$\,fF, and $E_J\approx85$\,GHz are extracted.

\section{Energy relaxation induced by quasiparticles tunneling}

\begin{figure}[hb]
    \centering
    \includegraphics{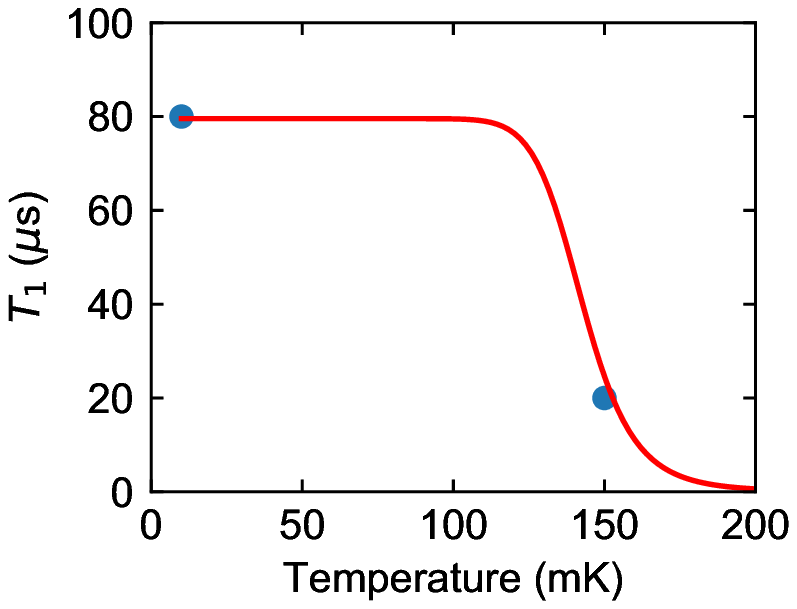}
    \caption{The temperature dependence of $T_1$ due to quasiparticle tunneling. The dots correspond to experimental data, and the red line represents the fitting results.}
    \label{fig2S}
\end{figure}

The temperature dependence of $T_1$ relaxation time due to quasiparticle tunneling effect is extensively described in the literature\,\cite{Catelani2011as, Catelani2011bs, Corcoles2011s, Paik2011s, Stern2014s}.
The relaxation rate between qubit states $n=0$ and $n=1$ is determined by the upward and downward transition rates:
\begin{equation}
    \Gamma = \Gamma_{1 \rightarrow 0} + \Gamma_{0 \rightarrow 1},
\end{equation}
where the transition rates are given be the sum of a non-equilibrium and a thermal equilibrium terms:
\begin{equation}
    \Gamma_{1 \rightarrow 0} = \Gamma_{eq, 1 \rightarrow 0} + \Gamma_{neq, 1 \rightarrow 0},
\end{equation}
and
\begin{equation}
    \Gamma_{0 \rightarrow 1} = \Gamma_{eq, 0 \rightarrow 1} + \Gamma_{neq, 0 \rightarrow 1}.
\end{equation}
For the thermal equilibrium term, it is assumed that the temperature is small compared to the superconducting gap, $k_B T \ll \Delta_0$ (for a 25-nm-thick aluminium film, $ \Delta_0 \approx 200$\,$\mu$eV), and the upward and downward transition rates are related by $ \Gamma_{eq, 0 \rightarrow 1} /  \Gamma_{eq, 1 \rightarrow 0}=e^{-\hbar \omega_{01}/ k_B T}$. As for the non-equlibrium part, the assumption is made that the quasiparticle energy (measured from the gap) is small compared to the qubit frequency $\delta E \ll \hbar \omega_{01}$ and therefore $\Gamma_{neq, 0 \rightarrow 1} \approx 0$.
Thus, the total relaxation rate is given by 
\begin{equation}
    \Gamma = \Gamma_{neq, 1 \rightarrow 0} + \Gamma_{eq, 1 \rightarrow 0} \left( 1 + e^{-\hbar \omega_{01}/ k_B T} \right).
\end{equation}
The relaxation rates $\Gamma_{neq, 1 \rightarrow 0}$ and $\Gamma_{eq, 1 \rightarrow 0}$ for multiple-junction qubits were derived in Ref.\,\cite{Catelani2011bs}:
\begin{equation}
\Gamma_{eq, 1 \rightarrow 0} = \sum_{j=1}^3 \left| \left<0|\sin \frac{\varphi_j}{2} | 1 \right> \right|^2 \frac{16 E_J^{(j)}}{\pi \hbar} e^{-\Delta_0/k_B T} e^{\hbar \omega_{01}/2 k_B T} \mathrm{K_0}\left( \frac{\hbar \omega_{01}}{2 k_B T}\right),
\end{equation}
\begin{equation}
    \Gamma_{neq, 1 \rightarrow 0} = \sum_{j=1}^3 \left| \left<0|\sin \frac{\varphi_j}{2} | 1 \right> \right|^2 x_{qp} \frac{8 E_J^{(j)}}{\pi \hbar} \sqrt{\frac{2\Delta_0}{\hbar \omega_{01}}},
\end{equation}
where $\varphi_j$ and $E_J^{(j)}$ are the phase and the Josephson energy of a $j$-th junction, $K_0()$ is the modified Bessel function of the second kind, and $x_{qp} = n_{qp}/2\nu_0 \Delta_0$ is the quasiparticle density normalized to the Cooper-pair density $n_{cp}\approx \nu_0 \Delta_0$ with $\nu_0$ being the density of states of conduction electrons at the Fermi level\,\cite{VanDuzer+Turner}. Since aluminium has a fcc crystal structure with a lattice constant of 4.05\,\AA, and a valence of 3, we estimate the normal-state conduction electron density to be $n\approx 18.8 \times  10^{22}$\,cm$^{-3}$. Consequently, since the density of states at the Fermi level is $\nu_0 = 3n/2E_F$ and the Fermi energy for aluminium is $E_F=11.6$\,eV, the value of Cooper-pair density $n_{cp}\approx 4.9 \times 10^6$\,$\mu$m$^{-3}$ is obtained. Thus, the total relaxation rate $\Gamma = T_1^{-1}$ can be written as
\begin{equation}
 \Gamma = A_{\sum} \left(\frac{8}{\pi}  x_{qp} \sqrt{\frac{2\Delta_0}{\hbar \omega_{01}}} + \frac{16}{\pi} e^{-\Delta_0/k_B T} e^{\hbar \omega_{01}/2 k_B T} \mathrm{K_0}\left( \frac{\hbar \omega_{01}}{2 k_B T}\right)\left( 1 + e^{-\hbar \omega_{01}/ k_B T} \right) \right),
    \label{eq:quas}
\end{equation}
where 
\begin{equation}
    A_{\sum} = \sum_{j=1}^3 \left| \left<0|\sin \frac{\varphi_j}{2} | 1 \right> \right|^2 E_J^{(j)} / \hbar.
\end{equation}
From the perturbation theory described in the Section \ref{section:perturbation}, the matrix elements can be estimated as:
\begin{equation}
    \left| \left< 0 \left|\sin \frac{\varphi_1}{2} \right| 1 \right> \right| \approx \left| \left< 0 \left| \sin \frac{\varphi_2}{2} \right| 1 \right> \right| \approx \left| \left< 0 \left|\sin \frac{\varphi_m}{2} \right| 1 \right> \right| = \frac{1}{2\sqrt{2}}\left( \frac{E_{C_S}}{E_J(1-2\alpha)}\right)^{1/4}\approx 0.13,
\end{equation}
and
\begin{equation}
    \left| \left< 0 \left|\sin \frac{\varphi_3}{2} \right| 1 \right> \right| \approx \left| \left< 0 \left|\cos \varphi_m \right| 1 \right> \right| = \frac{1}{4}\left( \frac{E_{C_S}}{E_J(1-2\alpha)}\right)^{1/2}\approx 0.03,
\end{equation}
using experimental values for the qubit parameters $\alpha=0.41$, $E_{Cs}=0.25$\,GHz, and $E_J=85$\,GHz.
Finally, the $T_1$ experimental data can be fitted using the equation (\ref{eq:quas}) with the single fitting parameter $x_{qp}$ as shown in Fig.\,\ref{fig2S}, from which an upper limit on the non-equilibrium quasiparticle density $n_{qp} \leq0.6$\,$\mu$m$^{-3}$ is determined.

\section{Dephasing due to thermal photon noise}
\begin{figure}[hb]
    \centering
    \includegraphics{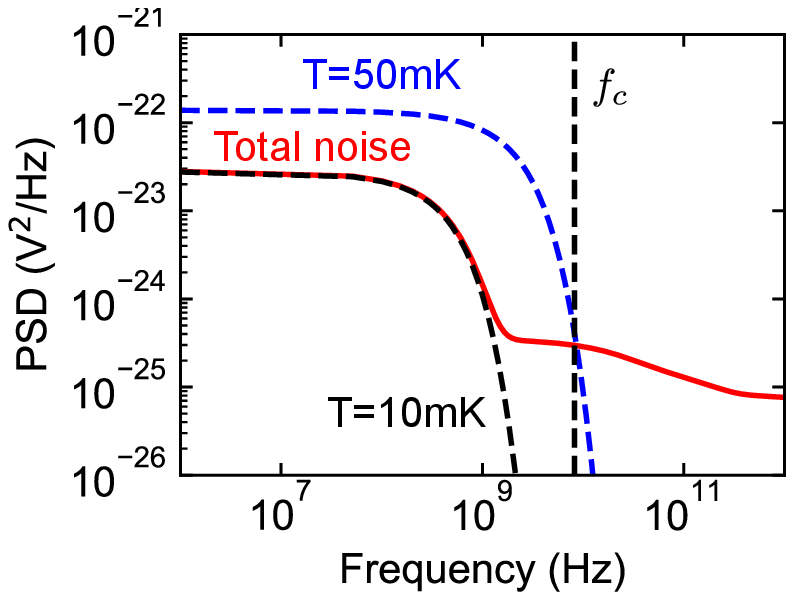}
    \caption{The effective temperature of cavity photons at the frequency $\omega_c=2 \pi f_c$ can be estimated graphically by comparing the total thermal noise (red line) with the thermal noise of the effective 50\,$\Omega$ load with the temperature $T_{\mathrm{eff}}$ (the black and blue curves correspond to $T_{\mathrm{eff}}=10$\,mK and $T_{\mathrm{eff}}=50$\,mK, respectively).}
    \label{fig3S}
\end{figure}
The dephasing rate due to the thermal cavity photon noise is given by\,\cite{Clerk2007s,Yan2016s}:
\begin{equation}
    \Gamma_{\phi}^{th} = \frac{\kappa^2}{\kappa^2+4\chi^2}\frac{4\chi^2}{\kappa} {\bar n},
    \label{eq:thermal-noise}
\end{equation}
where $\chi \approx 0.9$\,MHz and $\kappa \approx 1.3$\,MHz are the cavity pull and the cavity damping rate, respectively, as defined in the main text, and ${\bar n} = 1/(e^{\hbar \omega_c/k_B T_{\mathrm{eff}}}-1)$ is the thermal-photon population at the cavity frequency $\omega_c$. The effective temperature of the cavity photons $T_{\mathrm{eff}}$ can be estimated numerically from the condition\,\cite{Yan2018s}:
\begin{equation}
    S_{vv}(\omega_c,T_{\mathrm{eff}}) = \sum_i A_i S_{vv}(\omega_c,T_i),
    \label{eq:eff}
\end{equation}
where the function 
\begin{equation}
    S_{vv}(\omega,T) = 4 k_B T R \frac{\hbar \omega/k_B T}{e^{\hbar \omega/k_B T}-1}
\end{equation}
describes the power spectral density of the thermal noise from a $R=50$\,$\Omega$ resistor with a given temperature $T$, while $T_i$ are the temperatures of radiation sources (temperature stages of the dilution refrigerator), and coefficients $A_i$ are used to take into account total microwave attenuation between each source and the microwave cavity.
Equation\,(\ref{eq:eff}) can be solved graphically, as shown in Fig.\,\ref{fig3S}, and the effective temperature of thermal photons is estimated to be $T_{\mathrm{eff}}\approx 50$\,mK, which is equivalent to the average photon population of $\bar n \approx 0.0004$. Finally, using Eq.\,(\ref{eq:thermal-noise}), the dephasing time $T_\phi = \Gamma_\phi^{-1} \approx 3$\,ms is obtained.

\section{Filter function of the CPMG sequence}
\begin{figure}[hb]
    \centering
    \includegraphics{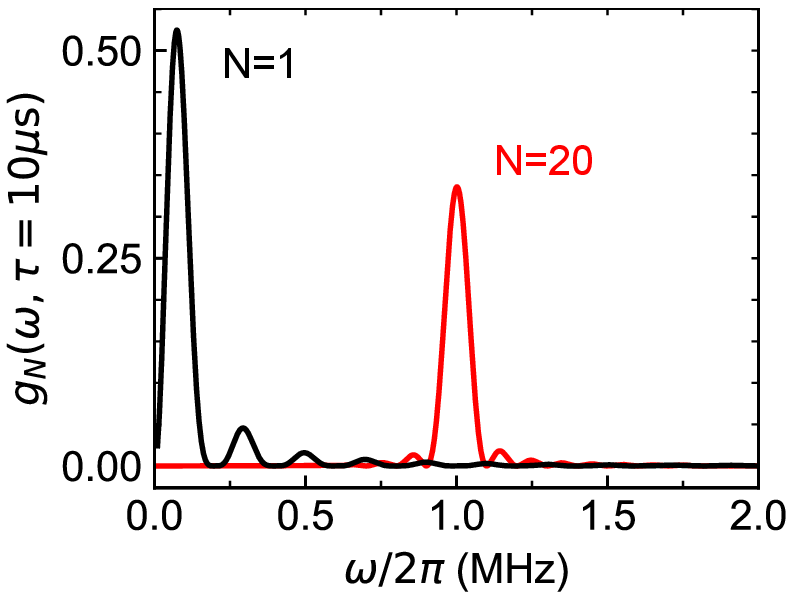}
    \caption{Filter functions for Hahn-echo and CPMG pulse sequences with $N=1$ and $N=20$, respectively.}
    \label{fig4S}
\end{figure}

The effect of a given CPMG pulse sequence on qubit dephasing can be explained as a frequency-domain filtering of the noise with the filter function given by\,\cite{Biercuk2009s, Bylander2011s}:
\begin{equation}
    g_N(\omega,\tau)=\frac{1}{(\omega \tau)^2} \left| 1 + (-1)^{N+1} e^{i\omega\tau} +2 \sum_{j=1}^N (-1)^j e^{i\omega \delta_j\tau} \cos{(\omega \tau_{\pi}/2)}\right|^2,
\end{equation}
where $\tau$ is the total length of the sequence, $N$ is the number of $\pi$ pulses, $\delta_j=t_j/\tau$ is the normalized position of the center of the $j$-th $\pi$-pulse of duration $\tau_{\pi}$. Filter functions for Hahn-echo and CPMG sequences used in the reported experiments are shown in Fig.\,\ref{fig4S}.

\providecommand{\noopsort}[1]{}\providecommand{\singleletter}[1]{#1}%


\begin{thebibliography}{28}
\bibitem [{not()}]{note1}%
  \BibitemOpen
  \href@noop {} {}\bibinfo {note} {Magnetic flux sensitivity was estimated from
  the experimental data at $\Phi\approx 0.502 \Phi_0$ using the equation
  $S_\Phi \approx |\partial \omega_{01} / \partial \Phi|^{-1} e/\sqrt{T_2}$
  which was derived for Ramsey-type
  magnetometry\,\cite{Danilin2018}.}\BibitemShut {Stop}%
\end{thebibliography}
\end{document}